\documentclass[12pt]{article}
\usepackage{amsmath}
\usepackage{epsfig}

\begin{document}
\begin{center}
\Large \bf Steppingstones in Hamiltonian dynamics \\

\end{center}

\begin{center}
Thomas F. Jordan\\
Physics Department \\
University of Minnesota\\
Duluth, Minnesota 55812\\
tjordan@d.umn.edu\\
\end{center}

\begin{displaymath}
\end{displaymath}
\begin{displaymath}
\end{displaymath}
\begin{displaymath}
\end{displaymath}
\begin{displaymath}
\end{displaymath}
\begin{displaymath}
\end{displaymath}
\begin{displaymath}
\end{displaymath}

       Easy steps through essential Hamiltonian dynamics are outlined, from necessary definitions of Poisson brackets and canonical transformations, to a quick proof that Hamiltonian evolution is made by canonical transformations, the quickest proof of Liouville's theorem, and on to Poincare-Cartan integral invariants and completely integrable dynamics, making room, providing tools, and setting the stage for more recent developments.

\pagebreak
\section*{I. Introduction}
\qquad At a few key points, my teaching of classical mechanics relies on calculations that I do differently from standard texts.  Strung together with bits of familiar material, they form an alternative route through Hamiltonian dynamics that I find advantageous.  It fits more easily into a course that moves from traditional to chaotic dynamics.  This requires that I provide notes for students.  Here I describe points that may be helpful for teachers as well.  They supplement what I have found in the standard texts.  I take care to refer to the texts I have found most helpful and those that are closest to what I do.

\qquad New developments in classical mechanics have given Hamiltonian dynamics a new role in our teaching.  It helps prepare us to work with the  methods and understand and appreciate the results of chaotic 
\begin{math}
\text{dynamics.}^{\mathbf{1-4}}
\end{math}
To make room for new developments, we have to delete or reduce some traditional topics.  Hamiltonian dynamics involves an accumulated wealth of mathematical development that can be shortened without sacrificing either the basic concepts or the results that illuminate contemporary work.  We do not need to know much about canonical transformations, for example, or anything about their generating functions, to work with the canonical transformations of Hamiltonian evolution.  Using the way they depend on time can be enough.

\qquad We can edit Hamiltonian dynamics for economy of expression and aim it in directions of current interest. We can, if we choose, step lightly over some traditional theory and save time for solving problems, or have time to introduce modern geometrical 
\begin{math}
\text{views.}^{\mathbf{1}}
\end{math}
Minimal steps through essential Hamiltonian dynamics are outlined here, from necessary definitions of Poisson brackets and canonical transformations, to a quick proof that Hamiltonian evolution is made by canonical transformations, the quickest proof of Liouville's theorem, and on toward contemporary goals, to Poincare-Cartan integral invariants, providing tools that can be used with surface-of-section 
\begin{math}
\text{maps,}^{\mathbf{5}}
\end{math}
and to completely integrable dynamics, setting the stage for use of action-angle variables and discussion of the effect of perturbations on resonant and non-resonant tori  and the KAM 
\begin{math}
\text{theorem.}^{\mathbf{6}}
\end{math}
Generating functions are particularly useful for proving that the transformation to action-angle variables is canonical, but that is proved here simply by calculating Poisson brackets.  We do not need a power tool for this one job; we can do it by hand.
\pagebreak

\section*{II. Poisson Brackets}
\qquad We work with canonical coordinates
\begin{math}
q_1,q_2, ... q_N
\end{math}
 and momenta 
\begin{math}
p_1, p_2, ... p_N.
\end{math}
 The Poisson bracket is defined by 

\begin{displaymath}
 \left[F,G \right] = \sum_{n = 1}^{N} \left( \frac{\partial F}{\partial q_n} \frac{\partial G}{\partial p_n} - \frac{\partial F}{\partial p_n} \frac{\partial G}{\partial q_n} \right)   \tag{2.1}
\end{displaymath}
for any two functions
\begin{math}
F
\end{math}
 and
\begin{math}
 G 
\end{math}
of the canonical coordinates and momenta.  It is linear for both
\begin{math}
F
\end{math}
 and
\begin{math}
 G
\end{math}
.  It is antisymmetric;
\begin{math}
\left[ G,F \right]
\end{math}
is
\begin{math}
-\left[ F,G \right]
\end{math}
.  These properties are obvious.  It is an easy exercise to prove that the Posisson bracket also satisfies the Jacobi identity
\begin{displaymath}
 \left[\left[F,G \right],H \right] + \left[\left[G,H \right],F \right] + \left[\left[H,F \right],G \right]  = 0. \tag{2.2}
\end{displaymath}
The canonical coordinates and momenta themselves have Poisson brackets
\begin{displaymath}
\left[q_m,q_n \right] = 0 \quad , \quad \left[p_m,p_n \right] = 0 
\end{displaymath}
\begin{displaymath}
\left[q_m,p_n \right] = \delta_{mn}  \tag{2.3}
\end{displaymath}
for m,n = 1,2, ... , N. Hamilton's equations
\begin{displaymath}
\frac{dq_n}{dt} = \frac{\partial H}{\partial p_n} \quad , \quad \frac{dp_n}{dt} = -\frac{\partial H}{\partial q_n} \tag{2.4}
\end{displaymath}
can be written as
\begin{displaymath}
 \frac{dq_n}{dt} = \left[q_n,H \right]  \quad , \quad \frac{dp_n}{dt} = \left[p_n,H \right].   \tag{2.5}
\end{displaymath}
Indeed, Hamilton's equations and the definition of the Poisson bracket imply that for any function 
\begin{math}
F
\end{math}
 of the canonical coordinates and momenta
\begin{displaymath}
\frac{dF}{dt} = \sum_{n = 1}^{N} \left( \frac{\partial F}{\partial q_n} \frac{\partial H}{\partial p_n} - \frac{\partial F}{\partial p_n} \frac{\partial H}{\partial q_n} \right) = \left[F,H \right]. \tag{2.6}
\end{displaymath}
The Hamiltonian
\begin{math}
H 
\end{math}
is a function of the canonical coordinates and momenta that may be different for different 
\begin{math}
t;
\end{math}
 it may be a function of the
\begin{math}
q_n,p_n 
\end{math}
and 
\begin{math}
t.
\end{math}

\pagebreak
\section*{III. Canonical Transformations}
\qquad The 2N-dimensional space of points specified by the canonical coordinates and momenta is called phase space. We consider changes of coordinates in phase space. Let 
\begin{math}
Q_n \text{,} P_n,
\end{math}
for n = 1,2,...,N, be functions of the 
\begin{math}
q_k \text{,} p_k
\end{math}
 that determine the 
\begin{math}
q_k \text{,} p_k
\end{math}
as functions of the 
\begin{math}
Q_n \text{,} P_n,
\end{math}
 so the 
\begin{math}
Q_n \text{,} P_n
\end{math}
 label the points of phase space as well as the 
\begin{math}
q_k \text{,} p_k
\end{math}
.  We say that the 
\begin{math}
Q_n \text{,} P_n
\end{math}
are canonical coordinates and momenta, and the transformation from the 
\begin{math}
q_k \text{,} p_k
\end{math}
to the 
\begin{math}
Q_n \text{,} P_n
\end{math}
 is a canonical transformation, if
\begin{displaymath}
\left[Q_m,Q_n \right] = 0 \quad , \quad \left[P_m,P_n \right] = 0 
\end{displaymath}
\begin{displaymath}
 \left[Q_m,P_n \right] = \delta_{mn}  \tag{3.1}
\end{displaymath}
for m,n = 1,2,...N.  Then the
\begin{math}
Q_n \text{,} P_n
\end{math}
can be used in place of the 
\begin{math}
q_n \text{,} p_n
\end{math}
 to calculate Poisson brackets; for any functions 
\begin{math}
F
\end{math}
 and 
\begin{math}
G
\end{math}
 of the phase-space variables,

\begin{displaymath}
\left[F,G \right] = \sum_{n,m} ( \frac{\partial F}{\partial Q_m} \left[Q_m,Q_n \right]  \frac{\partial G}{\partial Q_n} + \frac{\partial F}{\partial Q_m}\left[Q_m,P_n \right] \frac{\partial G}{\partial P_n} 
\end{displaymath}
\begin{displaymath}
+ \frac{\partial F}{\partial P_m} \left[P_m,Q_n \right]  \frac{\partial G}{\partial Q_n} +\frac{\partial F}{\partial P_m} \left[P_m,P_n \right]  \frac{\partial G}{\partial P_n} )
\end{displaymath}
\begin{displaymath}
   =   \sum_{n} \left( \frac{\partial F}{\partial Q_n} \frac{\partial G}{\partial P_n} - \frac{\partial F}{\partial P_n} \frac{\partial G}{\partial Q_n} \right).   \tag{3.2}
\end{displaymath}

From (2.6) and (3.2) we see that 
\begin{displaymath}
\frac{dQ_n}{dt} = \left[Q_n,H \right] = \frac{\partial H}{\partial P_n}
\end{displaymath}
\begin{displaymath}
 \frac{dP_n}{dt} = \left[P_n,H \right] = -\frac{\partial H}{\partial Q_n}.  \tag{3.3}
\end{displaymath}
Hamilton's equations are the same for the
\begin{math}
Q_n \text{,} P_n
\end{math}
as for the 
\begin{math}
q_n \text{,} p_n.
\end{math}
The Hamiltonian is the same thing written differently as a function of different variables.  Conversely, if this is true for any Hamiltonian, the transformation from the 
\begin{math}
q_k \text{,} p_k
\end{math}
 to the
\begin{math}
Q_n \text{,} P_n
\end{math}
 must be canonical; with (2.6) used to write the time derivatives as Poisson brackets, Hamilton's equations for the 
\begin{math}
Q_n \text{,} P_n
\end{math}
 give
\begin{displaymath}
 \left[Q_n,H \right] = \frac{\partial H}{\partial P_n} \text{\quad,\quad} \left[P_n,H \right] = -\frac{\partial H}{\partial Q_n}, \tag{3.4}
\end{displaymath}
which imply (3.1) in the particular cases where 
\begin{math}
H
\end{math}
 is 
\begin{math}
Q_m \text{ or } P_m.
\end{math}
\pagebreak

\section*{ IV. Hamiltonian Evolution}
\qquad When the dynamics is described by Hamilton's equations, the evolution in time is made by canonical transformations.  Let 
\begin{math}
Q_n \text{ and } P_n
\end{math}
be the functions of the 
\begin{math}
q_k \text{,} p_k
\end{math}
 and 
\begin{math}
t
\end{math}
 that are the solutions of the equations of motion
\begin{displaymath}
 \frac{dQ_n}{dt} = \left[Q_n,H \right]  \quad , \quad  \frac{dP_n}{dt} = \left[P_n,H \right]   \tag{4.1}
\end{displaymath}
specified by the boundary conditions
\begin{displaymath}
 Q_n(t = 0) = q_n  \quad ,  \quad P_n(t=0) = p_n.  \tag{4.2}
\end{displaymath}
The canonical coordinates and momenta
\begin{math}
q_n \text{ and } p_n
\end{math}
 at time zero evolve to 
\begin{math}
Q_n \text{,} P_n
\end{math}
at time 
\begin{math}
t.
\end{math}
  If their sets of values are
\begin{math}
\mathbf{q} \text{ and } \mathbf{p}
\end{math}
 at time zero, their values at time 
\begin{math}
t
\end{math}
 are 
\begin{math}
Q_n(\mathbf{q},\mathbf{p},t) \text{ and } P_n(\mathbf{q},\mathbf{p},t).
\end{math}

Explicitly, when the Hamiltonian does not depend on time and the series converges,
\begin{displaymath}
Q_n = q_n + t \left[q_n,H \right] + \frac{1}{2}t^2 \left[ \left[ q_n,H \right],H \right]
\end{displaymath}
\begin{displaymath}
 ... + \frac{1}{k!}t^k \left[ ... \left[q_n,H\right] ...,H \right] + ...  \tag{4.3}
\end{displaymath}

\noindent the bracket with 
\begin{math}
H
\end{math}
 being taken
\begin{math}
 k 
\end{math}
times in the term with
\begin{math}
t^{k}
\end{math}
; this and the corresponding formula for 
\begin{math}
P_n
\end{math}
do satisfy the equations of motion (4.1) and the boundary conditions (4.2).

The 
\begin{math}
Q_n \text{,} P_n
\end{math}
are canonical coordinates and momenta. The transformation from the 
\begin{math}
q_n \text{,} p_n
\end{math}
to the 
\begin{math}
Q_n \text{,} P_n
\end{math}
is a canonical transformation.  We prove this by showing that the 
\begin{math}
Q_n \text{,} P_n
\end{math}
satisfy the bracket relations (3.1).  Using the Jacobi identity (2.2) , we get
\begin{displaymath}
\frac{d}{dt} \left[Q_m,P_n\right] = \left[\left[Q_m,H\right],P_n\right] + \left[Q_m, \left[P_n,H\right]\right]
\end{displaymath}
\begin{displaymath}
= \left[\left[Q_m,P_n\right],H\right]. \tag{4.4}
\end{displaymath}
The Poisson bracket 
\begin{math}
\left[Q_m,P_n\right]
\end{math}
must be the function of the 
\begin{math}
q_k \text{,} p_k
\end{math}
and 
\begin{math}
t
\end{math}
 that is the solution of this equation of motion specified by the boundary condition
\begin{displaymath}
\left[Q_m,P_n\right](t = 0) = \left[q_m,q_n\right] = \delta_{mn},  \tag{4.5}
\end{displaymath}
which is 
\begin{math}
\left[Q_m,P_n\right] = \delta_{mn}.
\end{math}
Explicitly, for a time-independent Hamiltonian,
\begin{displaymath}
 \left[Q_m,P_n\right] = \left[q_m,p_n\right] + t \left[\left[q_m,p_n\right],H\right] + ...   = \delta_{mn}.   \tag{4.6}
\end{displaymath}
The other bracket relations (3.1) can be proved 
\begin{math}
\text{similarly.}^{\mathbf{7}}
\end{math}

\qquad The notation used here distinguishes the coordinates and momenta 
\begin{math}
Q_n, P_n
\end{math}
at time
\begin{math}
t
\end{math}
from the
\begin{math}
q_n, p_n
\end{math}
at time zero.  We will use this notation again when it is helpful, to consider the coordinates and momenta at time
\begin{math}
t
\end{math}
as functions of the initial values.  More often, since now we know that the coordinates and momenta are equally canonical at each time, we will follow the usual practice and work with canonical coordinates and momenta that depend on time, as we did in writing Hamilton's equations in Section II; this will be indicated by absence of
\begin{math}
Q_n , P_n.
\end{math}

\section*{V. Integral Invariants}

\qquad A point in phase space marks a set of values for the canonical coordinates and momenta. It moves through phase space as they change in time. The character of this motion reflects the structure of the dynamics.  Consider a set of points that occupy a volume in phase space.  If the dynamics is described by Hamilton's equations, the size of the volume the points occupy does not change in time as the points move, even though its shape generally does.  This property of Hamiltonian dynamics is called Liouville's theorem.

The proof is very 
\begin{math}
\text{simple.}^{\mathbf{8}}
\end{math}
At any time, the time derivative of the volume is the integral over the surface surrounding the volume of the normal component of the velocity of the motion of the points on the surface.  That is the integral over the volume of the divergence of the velocity, which is zero, because for Hamiltonian dynamics the divergence is

\begin{displaymath}
 \sum_{n = 1}^{N} \left( \frac{\partial}{\partial q_n} \frac{dq_n}{dt} + \frac{\partial}{\partial p_n} \frac{dp_n}{dt} \right) = \sum_{n = 1}^{N} \left( \frac{\partial^2 H}{\partial q_n\partial p_n} - \frac{\partial^2 H}{\partial p_n\partial q_n} \right) = 0.  \tag{5.1}
\end{displaymath}
We see that this holds for a time-dependent Hamiltonian; the Hamiltonian can be a different function of the 
\begin{math}
q_k , p_k
\end{math}
 at each time.

Finding quantities that do not change in time helps build a picture of the dynamics.  The volume of Liouville's theorem is one.  Others can be constructed as integrals along curves.

Consider a curve of points in phase space.  Let
\begin{math}
 \theta
\end{math}
 be a parameter that measures location along the curve.  The phase-space coordinates 
\begin{math}
q_n, p_n
\end{math}
of points on the curve are functions of
\begin{math}
 \theta.
\end{math}
  As a point on the curve moves in time, its phase-space coordinates change from 
\begin{math}
q_n(\theta) , p_n(\theta) 
\end{math}
 at time zero to 
\begin{displaymath}
Q_n ( \mathbf{q} ( \theta ) , \mathbf{p} ( \theta ) , t )  \qquad , \qquad P_n ( \mathbf{q} ( \theta ) , \mathbf{p} ( \theta ) , t ) \tag{5.2}
\end{displaymath}
at time
\begin{math}
t,
\end{math}
where 
\begin{math}
Q_n \text{ and } P_n
\end{math}
are the solutions of the equations of motion (4.1) that satisfy the boundary conditions (4.2).  We write 
\begin{math}
\mathbf{q} \text{ for the set of } q_k 
\end{math}
and
\begin{math}
\mathbf{p} \text{ for the set of } p_k.
\end{math}
The phase-space coordinates of the moving point are functions of
\begin{math}
\theta \text{ and } t.
\end{math}
We can let points at different
\begin{math}
\theta
\end{math}
 move for different intervals of time
\begin{math}
t.
\end{math}
  If we make the time interval zero at the ends of a curve segment, the moved segment and the original segment will form a closed loop, as shown in Figure 1.  We can replace the original segment with another moved curve to make a closed loop on which the phase-space coordinates change as functions of both 
\begin{math}
 \theta \text{ and } t
\end{math}
 all the way around.  In this way we can generate any closed loop in phase space that surrounds an area of surface of tracks of motion in time, an area of surface swept by a curve segment moving in time.

For any such loop,
\begin{displaymath}
 \oint ( \sum_{n = 1}^N P_n dQ_n - H dt) = 0 ,  \tag{5.3}
\end{displaymath}
the integral being around the closed loop. We can see this from
\begin{displaymath}
\oint ( \sum_{n = 1}^N P_n (\frac{\partial Q_n}{\partial \theta} d\theta + \frac{\partial Q_n}{\partial t} dt ) - H dt)
\end{displaymath}
\begin{displaymath}
 = \oint  ( A_\theta d\theta + A_t dt) = \oint \vec{A} \cdot d\vec{l} = \int \nabla \times \vec{A} \cdot d\vec{S}  \tag{5.4}
\end{displaymath}
with
\begin{displaymath}
 A_\theta = \sum_{n = 1}^N P_n \frac{\partial Q_n}{\partial \theta} \qquad , \qquad A_t = \sum_{n = 1}^{N} P_n \frac{\partial Q_n}{\partial t} - H,  \tag{5.5}
\end{displaymath}
the integrals involving
\begin{math}
 \vec{A}
\end{math}
 being around the closed loop in the
\begin{math}
\theta , t
\end{math}
 plane and over the area it encloses; because for Hamiltonian dynamics
\begin{displaymath}
\frac{\partial A_\theta}{\partial t} - \frac{\partial A_t}{\partial \theta} = \sum_{n = 1}^N \frac{\partial P_n}{\partial t}  \frac{\partial Q_n}{\partial \theta} - \sum_{n = 1}^N \frac{\partial P_n}{\partial \theta}  \frac{\partial Q_n}{\partial t} + \frac{\partial H}{\partial \theta} 
\end{displaymath}
\begin{displaymath}
 = \sum_{n = 1}^N ( - \frac{\partial H}{\partial Q_n}\frac{\partial Q_n}{\partial \theta}) -   \sum_{n = 1}^N  \frac{\partial P_n}{\partial \theta}\frac{\partial H}{\partial P_n} +  \frac{\partial H}{\partial \theta}  = 0. \tag{5.6}
\end{displaymath}

This implies that when a closed loop in phase space is changed by letting the points on it move for different intervals of time, the integral around the loop
\begin{displaymath}
\oint ( \sum_{n= 1}^N P_n dQ_n - H dt)  \tag{5.7}
\end{displaymath}
is not changed.  We can see this by letting our original curve be a closed loop and letting the points on it move for different intervals of time to form a moved loop.  Connecting the two loops along two closely adjacent tracks of motion, as in Figure 2, yields a closed loop to which (5.3) applies.  It surrounds a surface of tracks of motion, a surface swept by a curve moving in time.  Since the integrals in opposite directions along the two tracks of motion cancel, the integrals in opposite directions around the two loops must cancel, which means the integrals in the same direction around the two loops are the same.

These results are important for surface-of-section 
\begin{math}
\text{maps.}^{\mathbf{5}}
\end{math}
The quantities (5.7) that do not change in time are called Poincare-Cartan integral 
\begin{math}
\text{invariants.}^{\mathbf{9}}
\end{math}
When the 
\begin{math}
H dt
\end{math}
 term is absent, they are called Poincare integral 
\begin{math}
\text{invariants.}^{\mathbf{10}}
\end{math}
The 
\begin{math}
H dt
\end{math}
 term will be absent if there are no time differentials or if
\begin{math}
 H 
\end{math} can be taken outside the integral.  We will use (5.3) in the next section.

\pagebreak
\section*{VI. Complete Integrability}
\qquad A canonical transformation can make the equations of motion so simple that their solutions become trivial.  This is possible if the dynamics are completely integrable.  That means there are N constants of the motion, which we call 
\begin{math}
I_n
\end{math}
with 
\begin{math}
n
\end{math}
 running from 1 to N, that in addition to
\begin{displaymath}
\left[ I_n , H \right] = 0 \tag{6.1}
\end{displaymath}
which says they are constants, satisfy
\begin{displaymath}
\left[ I_m , I_n \right] = 0 \tag{6.2}
\end{displaymath}
for 
\begin{math}
m,n = 1,2,...N.
\end{math}
 It is assumed also that these
\begin{math}
I_n
\end{math}
are functions of the 
\begin{math}
q_k , p_k \text{ that determine the } p_k
\end{math}
as functions of the 
\begin{math}
q_n , I_n
\end{math}
 in a sufficiently relevant region of phase space, so the 
\begin{math}
q_n , I_n
\end{math}
 label those points of phase space as well as the 
\begin{math}
q_k , p_k.
\end{math}

These assumptions of complete integrability imply, as we will show, that there are 
\begin{math}
Q_n \text{ for which the }
Q_n,I_n
\end{math}
are canonical coordinates and 
\begin{math}
\text{momenta.}^{\mathbf{11}}
\end{math}
The
\begin{math}
P_n \text{ are the } I_n.
\end{math}
They are all constants of the motion.  That means Hamilton's equations give
\begin{displaymath}
\frac{\partial H}{\partial Q_n} = - \frac{dP_n}{dt} = 0. \tag{6.3}
\end{displaymath}
Then 
\begin{math}
H
\end{math}
 is a function only of the 
\begin{math}
P_n 
\end{math}
 and
\begin{displaymath}
\frac{dQ_n}{dt} = \frac{\partial H}{\partial P_n} \tag{6.4}
\end{displaymath}
is a constant of the motion.  The equations of motion are solved.  The 
\begin{math}
P_n
\end{math}
 are constant in time.  The 
\begin{math}
Q_n
\end{math}
 change at constant rates; they are linear functions of time.  Action-angle variables are a particular case of these canonical coordinates and momenta
\begin{math}
Q_n \text{ and } I_n
\end{math}
 for completely integrable dynamics.

We will take two steps to prove that there are canonical coordinate partners
\begin{math}
Q_n
\end{math}
 for canonical momenta 
\begin{math}
I_n.
\end{math}
The first step, needed to define the 
\begin{math}
Q_n,
\end{math}
 is to show that there is a function 
\begin{math}
S
\end{math}
 of the 
\begin{math}
q_n , I_n
\end{math}
such that 
\begin{displaymath}
\frac{\partial S}{\partial q_n} = p_n. \tag{6.5}
\end{displaymath}
It is an action integral. Let
\begin{displaymath}
S(\mathbf{q},\mathbf{I}) = \int_{\mathbf{q_0}}^{\mathbf{q}} \sum_{k = 1}^N p_k(\mathbf{q'} ,\mathbf{ I} ) dq'_k. \tag{6.6}
\end{displaymath}
We write 
\begin{math}
\mathbf{q}
\end{math}
for the set of 
\begin{math}
q_n
\end{math}
and
\begin{math}
\mathbf{I}
\end{math}
for the set of 
\begin{math}
I_n.
\end{math}
For each fixed set of values of the 
\begin{math}
I_n,
\end{math}
the integral gives
\begin{math}
S
\end{math}
as a function of the 
\begin{math}
q_n.
\end{math}
The integral is along a path on the surface in phase space where the 
\begin{math}
I_n
\end{math}
are constant, starting at the point where the 
\begin{math}
\mathbf{q}
\end{math}
have values 
\begin{math}
\mathbf{q_0}.
\end{math}
  To establish that
\begin{math}
S
\end{math}
 is well defined, we need to show that the integral is not changed by variations in the path.

We have described the motion in phase space generated by the Hamiltonian
\begin{math}
H
\end{math}
 for evolution in time.  Now we consider also the motion that is generated the same way when the Hamiltonian is replaced by one of the 
\begin{math}
I_n
\end{math}
or a linear combination of the
\begin{math}
I_n.
\end{math}
From the bracket relations (6.2) we see that this motion does not change any of the 
\begin{math}
I_m.
\end{math}
  It stays on the surface in phase space where the 
\begin{math}
I_m
\end{math}
stay constant.  The velocity vector for the motion generated by 
\begin{math}
I_n,
\end{math}
 the vector that points along the direction of the motion in phase space and measures the rate of the motion with respect to a parameter 
\begin{math}
s
\end{math}
along the curve of the motion, has components
\begin{displaymath}
\frac{dq_1}{ds} = \frac{\partial I_n}{\partial p_1} , \qquad...\qquad \frac{dq_N}{ds} = \frac{\partial I_n}{\partial p_N}.
\end{displaymath}
\begin{displaymath}
\frac{dp_1}{ds} = - \frac{\partial I_n}{\partial q_1} ,\qquad ...\qquad \frac{dp_N}{ds} = - \frac{\partial I_n}{\partial q_N}. \tag{6.7}
\end{displaymath}
We are assuming that the 
\begin{math}
q_n , I_n
\end{math}
are coordinates for a region of phase space.  For each fixed set of values of the 
\begin{math}
q_n ,\text{ the } I_n
\end{math}
must be coordinates for an N-dimensional space.  They must vary in N different directions.  Their gradients must be N linearly independent vectors.  From (6.7), which relates the components of the gradient of 
\begin{math}
 I_n
\end{math}
to the components of the velocity vector for the motion generated by 
\begin{math}
 I_n,
\end{math}
we see that the N velocity vectors for the motions generated by the 
\begin{math}
I_n
\end{math}
 must be linearly independent.  The velocity vectors for the motions generated by the 
\begin{math}
 I_n
\end{math}
and linear combinations of the 
\begin{math}
 I_n
\end{math}
point in all directions on the surface where the
\begin{math}
I_m
\end{math}
are constant.

Now we can prove that the integral (6.6) is not changed by variations in the path on the surface of constant 
\begin{math}
 I_m.
\end{math}
  If an infinitesimal segment of the path is moved on the surface of constant
\begin{math}
 I_m,
\end{math}
the moved segment and the original segment form a closed loop, as in Figure 1. Now the motion is generated by a linear combination of the 
\begin{math}
I_n
\end{math}
instead of
\begin{math}
H.
\end{math}
We can use (5.3) with this replacement.  The 
\begin{math}
H dt
\end{math}
 term becomes zero, because the linear combination of the 
\begin{math}
I_n
\end{math}
 is a constant that can be taken out of the integral, and the change in the parameter 
\begin{math}
t
\end{math}
or 
\begin{math}
s
\end{math}
around the closed loop is zero.  Therefore
\begin{displaymath}
\oint \sum_{k = 1}^N P_k dQ_k = 0 \tag{6.8}
\end{displaymath}
for the closed loop, which means the integral (6.6) is the same for the new path as for the old.  We conclude that 
\begin{math}
S
\end{math}
is well defined.  It clearly satisfies (6.5).

That allows us to define the 
\begin{math}
Q_n.
\end{math}
Let
\begin{displaymath}
Q_n = \frac{\partial S}{\partial I_n}. \tag{6.9}
\end{displaymath}
The final step is to show that the 
\begin{math}
Q_n
\end{math}
and 
\begin{math}
 I_n
\end{math}
are canonical coordinates and momenta, that the bracket relations (3.1) are satisfied with 
\begin{math}
 I_n
\end{math}
for 
\begin{math}
P_n
\end{math}
 and these
\begin{math}
Q_n.
\end{math}

Using (6.2), (6.5) and (6.9) and taking partial derivatives of 
\begin{math}
S,
\end{math}
the 
\begin{math}
Q_n
\end{math}
 and the 
\begin{math}
p_n
\end{math}
with respect to the 
\begin{math}
q_k , I_k
\end{math}
and partial derivatives of the 
\begin{math}
 I_n
\end{math}
with respect to the 
\begin{math}
q_k , p_k,
\end{math}
we get
\begin{displaymath}
\left[ Q_m, I_n \right] = \sum_{j = 1}^N \frac{\partial Q_m}{\partial q_j} \left[ q_j , I_n \right] + \sum_{j = 1}^N \frac{\partial Q_m}{\partial I_j} \left[ I_j , I_n \right]
\end{displaymath}
\begin{displaymath}
= \sum_{j = 1}^N \frac{\partial^2 S}{\partial q_j \partial I_m} \frac{\partial I_n}{\partial p_j} = \sum_{j = 1}^N \frac{\partial p_j}{\partial I_m} \frac{\partial I_n}{\partial p_j}
\end{displaymath}
\begin{displaymath}
= \sum_{j = 1}^N \frac{\partial I_n}{\partial p_j} \frac{\partial p_j}{\partial I_m} + \sum_{j = 1}^N \frac{\partial I_n}{\partial q_j} \frac{\partial q_j}{\partial I_m}
\end{displaymath}
\begin{displaymath}
= \frac{\partial I_n}{\partial I_m} = \delta_{mn};  \tag{6.10}
\end{displaymath}
in this context 
\begin{math}
\partial q_j / \partial I_m
\end{math}
 is zero because the derivatives of 
\begin{math}
 I_n
\end{math}
 are with respect to the
\begin{math}
q_j
\end{math}
and
\begin{math}
p_j
\end{math}
and the derivatives of the
\begin{math}
p_j
\end{math}
and
\begin{math}
q_j
\end{math}
 are with respect to the 
\begin{math}
q_n
\end{math}
and 
\begin{math}
 I_m.
\end{math}
Similarly, we get
\pagebreak

\begin{displaymath}
\left[ Q_m ,Q_n \right] = \sum_{i,j = 1}^N \frac{\partial Q_m}{\partial q_i} \left[ q_i , q_j \right] \frac{\partial Q_n}{\partial q_j}
\end{displaymath}
\begin{displaymath}
+ \sum_{i,j=1}^N \frac{\partial Q_m}{\partial I_i} \left[ I_i , q_j \right] \frac{\partial Q_n}{\partial q_j} + \sum_{j =1}^N \left[ Q_m , I_j \right]\frac{\partial Q_n}{\partial I_j}
\end{displaymath}
\begin{displaymath}
= \sum_{i,j = 1}^N \frac{\partial Q_m}{\partial I_i} ( - \frac{\partial I_i}{\partial p_j} ) \frac{\partial Q_n}{\partial q_j} + \frac{\partial Q_n}{\partial I_m}
\end{displaymath}
\begin{displaymath}
= - \sum_{i,j = 1}^N \frac{\partial^2 S}{\partial I_i \partial I_m} \frac{\partial I_i}{\partial p_j} \frac{\partial^2 S}{\partial q_j \partial I_n} + \frac{\partial^2 S}{\partial I_m \partial I_n}
\end{displaymath}
\begin{displaymath}
= - \sum_{i,j = 1}^N \frac{\partial^2 S}{\partial I_i \partial I_m} \frac{\partial I_i}{\partial p_j} \frac{\partial p_j}{\partial I_n} + \frac{\partial^2 S}{\partial I_m \partial I_n}
\end{displaymath}
\begin{displaymath}
= - \frac{\partial^2 S}{\partial I_n \partial I_m} + \frac{\partial^2 S}{\partial I_m \partial I_n} = 0 \tag{6.11}
\end{displaymath}
by using again the last steps of (6.10) which show that in this context
\begin{displaymath}
\sum_{j} \frac{\partial I_i}{\partial p_j} \frac{\partial p_j}{\partial I_m} = \delta_{im}. \tag{6.12}
\end{displaymath}

\pagebreak
\noindent \begin{math}
^{\mathbf{1}}
\end{math}
S. N. Rasband, \emph{Dynamics} (Wiley, New York, 1983).\\

\noindent \begin{math}
^{\mathbf{2}}
\end{math}
M. Tabor, \emph{Chaos and integrability in nonlinear dynamics} (Wiley, New York, 1989).\\

\noindent \begin{math}
^{\mathbf{3}}
\end{math}
E. Ott, \emph{Chaos in dynamical systems} (Cambridge U. Press, Cambridge, 1993).\\

\noindent \begin{math}
^{\mathbf{4}}
\end{math}
H. Goldstein, C. Poole and J. Safko, \emph{Classical Mechanics} (Addison Wesley, San Francisco, 2002).\\

\noindent \begin{math}
^{\mathbf{5}}
\end{math}
Reference 2, pp. 123-6 and 183-4.\\

\noindent \begin{math}
^{\mathbf{6}}
\end{math}
Reference 1, Chapter 9; reference 2, Chapters 3, 4; reference 3, Chapter 7.\\

\noindent \begin{math}
^{\mathbf{7}}
\end{math}
That Hamiltonian evolutions are canonical transformations is proved with generating functions by J. L. Synge and B.A. Griffith, \emph{ Priciples of Mechanics} (McGraw-Hill, New York, 1959), pp. 473-4; and H.C. Corben and P. Stehle, \emph{Classical Mechanics} (Wiley, New York, 1960), pp. 215-7.\\

\noindent \begin{math}
^{\mathbf{8}}
\end{math}
Reference 3, pp. 209-10; J.L. Synge and B.A. Griffith, reference 7, pp. 476-7 ; J.R. Taylor, \emph{Classical Mechanics} (University Science Books, New York, 2002), Section 13.7.  It is not necessary to use a Jacobian or to consider a fluid or a density.\\

\noindent \begin{math}
^{\mathbf{9}}
\end{math}
Poincare-Cartan integral invariants are usually discussed in the context of an extended phase space that includes the time dimension.  For example, see reference 3, pp. 212-3.\\

\noindent \begin{math}
^{\mathbf{10}}
\end{math}
Poincare integral invariants are established very simply with generating functions by H.C. Corben and P. Stehle, reference 7, pp. 236-7. An integral around the curve is used by J.L. Synge and B. A. Griffith, reference 7, p. 473; and L.A. Pars, \emph{A Treatise on Analytical Dynamics} (Ox Bow, Woodridge, Connecticut, 1979), p. 433.\\

\noindent \begin{math}
^{\mathbf{11}}
\end{math}
The method of proof used here is outlined in reference 1, pp. 156-8. For the usual method using generating functions see, for example, reference 4; H.C. Corben and P. Shehle, reference 7; or L.A. Pars, reference 10.
\pagebreak 
\begin{figure}
\begin{center}
\epsfig{file=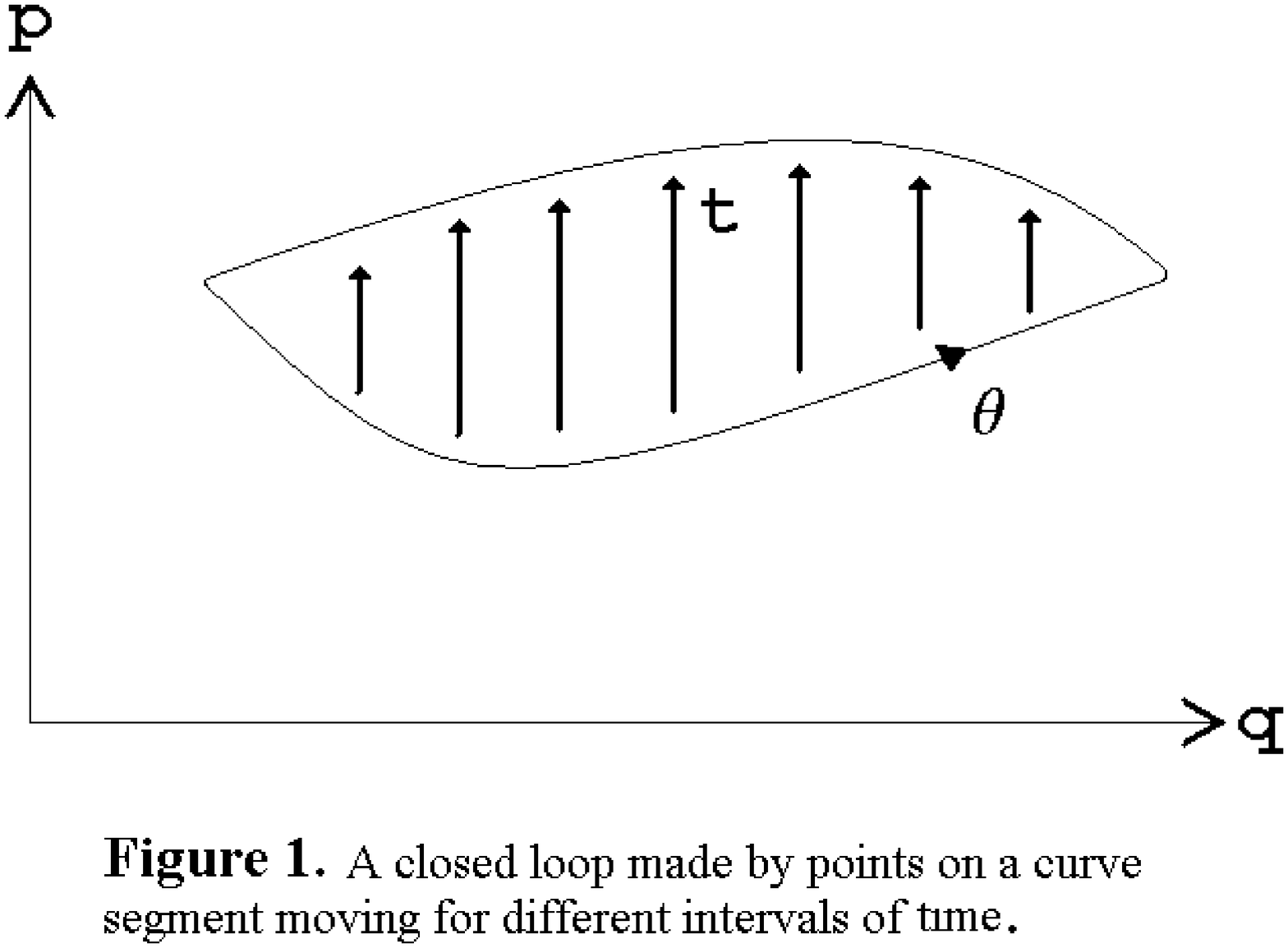, width=12cm}
\end{center}
\end{figure}

\pagebreak

\begin{figure}
\begin{center}
\epsfig{file=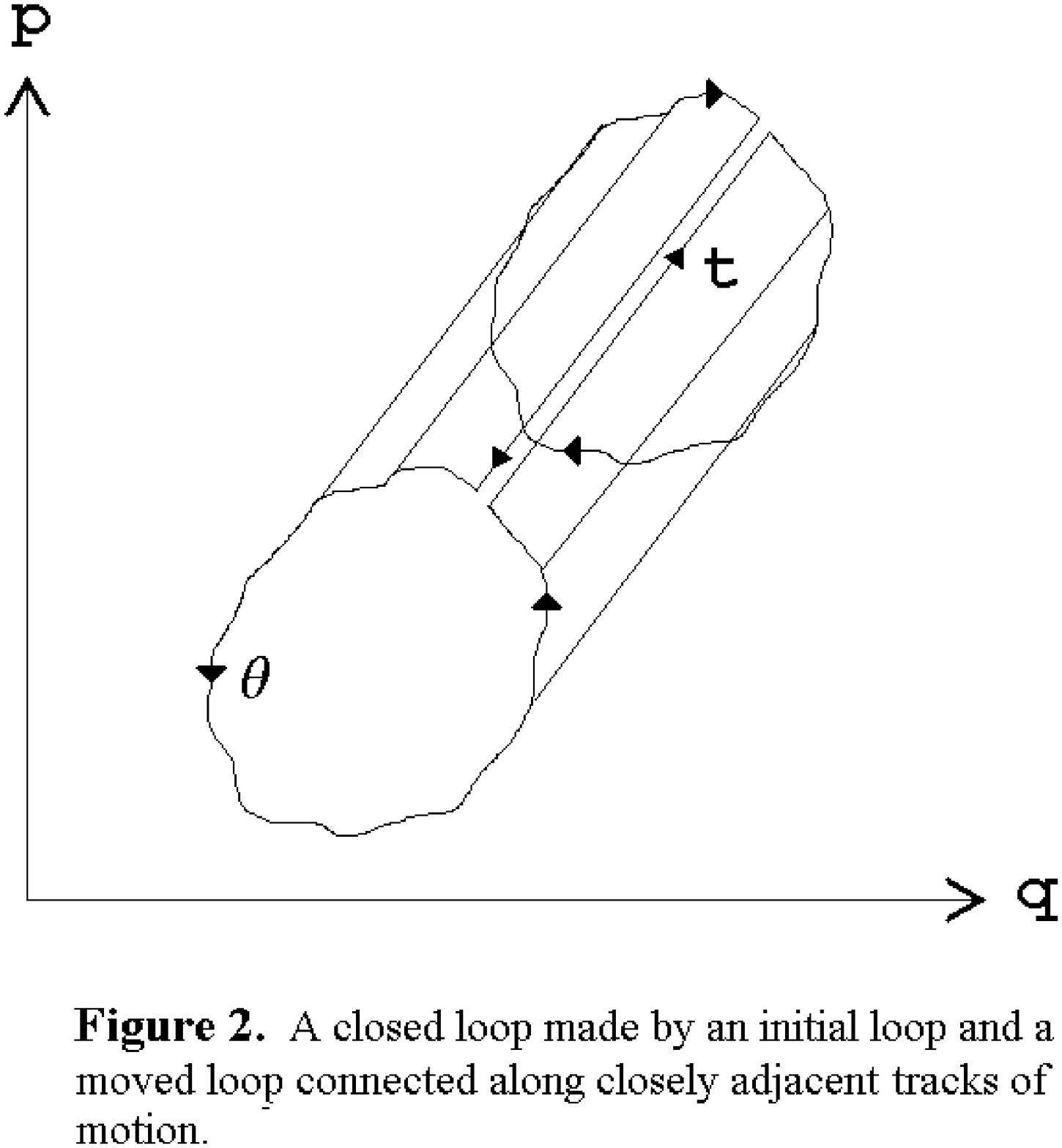, width=12cm}
\end{center}
\end{figure}

\end{document}